\documentclass[letterpaper, 10 pt, conference]{ieeeconf}  

\IEEEoverridecommandlockouts                              
\usepackage{epsfig} 
\usepackage{amsmath} 
\usepackage{amssymb}  

\usepackage{theorem}
\usepackage{stmaryrd}
\usepackage{color}
\usepackage{bm}

\IEEEtriggeratref{25}

\title{\LARGE \bf
LP Decoding meets LP Decoding: \\
A Connection between Channel Coding and Compressed Sensing${}^{*}$
}

\author{Alexandros G.~Dimakis
        and
        Pascal O.~Vontobel
\thanks{A.~G.~Dimakis is with the
        Department of Electrical Engineering-Systems, 
        University of Southern California,
        Los Angeles, CA 90089, USA
        (email: dimakis@usc.edu).}%
\thanks{P.~O.~Vontobel is with
        Hewlett--Packard Laboratories,
        1501 Page Mill Road,
        Palo Alto, CA 94304, USA
        (email: pascal.vontobel@ieee.org).}%
\thanks{${}^{*}$Appeared in the Proceedings of the 
  47th Allerton Conference on
  Communications, Control, and Computing, Allerton House, Monticello, 
  Illinois, USA, Sep.~30--Oct.~2, 2009. This version of the paper 
  contains all 
  the proofs that were omitted in the official version
  due to space 
  limitations.}%
}

\renewcommand{\mathbf}[1]{{\bm{#1}}}     

\newcommand{\ignore}[1]{}

\newcommand{\supp}{\operatorname{supp}}

\newcommand{\dint}[1]{\operatorname{d}{#1}}

\newcommand{\matr}[1]{\mathbf{#1}}
\newcommand{\vect}[1]{\mathbf{#1}}
\newcommand{\code}[1]{\mathcal{#1}}

\newcommand{\set}[1]{\mathcal{#1}}

\newcommand{\GF}[1]{\mathbb{F}_{#1}}
\newcommand{\R}{\mathbb{R}}

\newcommand{\Rp}{\mathbb{R}_{\geq 0}}
\newcommand{\Rpp}{\mathbb{R}_{> 0}}

\newcommand{\tr}{\mathsf{T}}

\newcommand{\codeCQC}[1]{\code{C}_{\mathrm{QC}}^{(r)}}

\newcommand{\defeq}{\triangleq}

\newcommand{\vnu}{\boldsymbol{\nu}}
\newcommand{\vnuS}{\vnu_{\setS}}
\newcommand{\vnuSc}{\vnu_{\setSc}}
\newcommand{\vnnu}{\vnu}

\newcommand{\vnnuS}{\vnu_{\setS}}
\newcommand{\vnnuSc}{\vnu_{\setSc}}
\newcommand{\vlambda}{\boldsymbol{\lambda}}
\newcommand{\vomega}{\boldsymbol{\omega}}

\newcommand{\convhull}{\operatorname{conv}}
\newcommand{\conichull}{\operatorname{conic}}

\newcommand{\zeronorm}[1]{\lVert #1 \rVert_0}
\newcommand{\onenorm}[1]{\lVert #1 \rVert_1}
\newcommand{\twonorm}[1]{\lVert #1 \rVert_2}
\newcommand{\infnorm}[1]{\lVert #1 \rVert_{\infty}}
\newcommand{\norm}[2]{\lVert #1 \rVert_{#2}}

\newcommand{\vzero}{\vect{0}}
\newcommand{\va}{\vect{a}}

\newcommand{\vb}{\vect{b}}

\newcommand{\ve}{\vect{e}}

\newcommand{\veS}{\vect{e}_{\setS}}
\newcommand{\veSc}{\vect{e}_{\setSc}}
\newcommand{\hve}{\vect{\hat e}}

\newcommand{\vs}{\vect{s}}

\newcommand{\vu}{\vect{u}}
\newcommand{\hvu}{\vect{\hat u}}

\newcommand{\vX}{\vect{X}}
\newcommand{\vY}{\vect{Y}}
\newcommand{\vx}{\vect{x}}

\newcommand{\vy}{\vect{y}}

\newcommand{\Z}{\mathbb{Z}}
\newcommand{\Zp}{\Z_{\geq 0}}
\newcommand{\Zpp}{\Z_{> 0}}

\newcommand{\hvx}{\vect{\hat x}}

\renewcommand{\leq}{\leqslant}
\renewcommand{\geq}{\geqslant}

\newcommand{\setI}{\set{I}}
\newcommand{\setJ}{\set{J}}

\newcommand{\setS}{\set{S}}
\newcommand{\setSc}{\overline{\setS}}

\newcommand{\nullspaceR}{\operatorname{nullspace}_{\R}}
\newcommand{\nullspaceGFtwo}{\operatorname{nullspace}_{\GF{2}}}
\newcommand{\NSPR}{\mathrm{NSP}^{\leq}_{\R}}
\newcommand{\SNSPR}{\mathrm{NSP}^{<}_{\R}}
\newcommand{\setweightR}[2]{\Sigma_{\R^{#1}}^{(#2)}} 
\newcommand{\setweightGFtwo}[2]{\Sigma_{\GF{2}^{#1}}^{(#2)}} 

\newtheorem{lemma}{Lemma}
\newtheorem{theorem}[lemma]{Theorem}

\theoremstyle{plain}

\newtheorem{PreDefinition}[lemma]{{\textbf{Definition}}}
  \newenvironment{definition}%
    {\begin{PreDefinition}}{\hfill$\square$\end{PreDefinition}}

\theoremstyle{plain}

\newtheorem{PreRemark}[lemma]{{\textbf{Remark}}}
    {\begin{PreRemark}\upshape}{\hfill$\square$\end{PreRemark}}

\newtheorem{PreExample}[lemma]{{\textbf{Example}}}
    {\begin{PreExample}\upshape}{\hfill$\square$\end{PreExample}}

\newcommand{\rank}{\operatorname{rank}}
\newcommand{\fp}[1]{\set{#1}}

\newcommand{\fc}[1]{\set{#1}}

\newcommand{\wpsAWGNC}{w_{\mathrm{p}}^{\mathrm{AWGNC}}}
\newcommand{\wpsBSC}{w_{\mathrm{p}}^{\mathrm{BSC}}}
\newcommand{\wpsBSCmod}{w_{\mathrm{p}}^{\mathrm{BSC}'}}
\newcommand{\wpsBEC}{w_{\mathrm{p}}^{\mathrm{BEC}}}

\newcommand{\wpsAWGNCmin}{w_{\mathrm{p}}^{\mathrm{AWGNC,min}}}
\newcommand{\wpsBSCmin}{w_{\mathrm{p}}^{\mathrm{BSC,min}}}
\newcommand{\wpsBECmin}{w_{\mathrm{p}}^{\mathrm{BEC,min}}}

\newcommand{\wmaxfr}{w_{\mathrm{max-frac}}}
\newcommand{\wmaxfrmin}{w^{\mathrm{min}}_{\mathrm{max-frac}}}

\newcommand{\card}[1]{\# #1}

\newcommand{\wH}{w_{\mathrm{H}}}

\newcommand{\codeCCC}{\code{C}_{\mathrm{CC}}}
\newcommand{\matrGCC}{\matr{G}_{\mathrm{CC}}}
\newcommand{\matrHCC}{\matr{H}_{\mathrm{CC}}}
\newcommand{\setXCC}{\set{X}_{\mathrm{CC}}}
\newcommand{\setYCC}{\set{Y}_{\mathrm{CC}}}
\newcommand{\codeCCCdown}[1]{\code{C}_{\mathrm{CC},#1}}

\newcommand{\codeCCS}{\code{C}_{\mathrm{CS}}}
\newcommand{\matrGCS}{\matr{G}_{\mathrm{CS}}}
\newcommand{\matrHCS}{\matr{H}_{\mathrm{CS}}}
\newcommand{\setXCS}{\set{X}_{\mathrm{CS}}}
\newcommand{\setYCS}{\set{Y}_{\mathrm{CS}}}

\newcommand{\inGFtwo}{\text{(in $\GF{2}$)}}

\newcommand{\CSOPT}{\textbf{CS-OPT}}
\newcommand{\CSLPD}{\textbf{CS-LPD}}

\newcommand{\CCMLD}{\textbf{CC-MLD}}
\newcommand{\CCMLDone}{\textbf{CC-MLD1}}
\newcommand{\CCMLDtwo}{\textbf{CC-MLD2}}
\newcommand{\CCMLDthree}{\textbf{CC-MLD3}}
\newcommand{\CCLPD}{\textbf{CC-LPD}}

\newcommand{\optprog}[2]
{%
  \noindent\mbox{}\\[0cm]
  \noindent\fbox{%
  \begin{minipage}{0.955\linewidth}
    \mbox{}\\[-0.5cm]
    #1\\[#2]
  \end{minipage}
  }
  \noindent\mbox{}\\[-0.2cm]
}

\begin{document}

\maketitle
\thispagestyle{empty}
\pagestyle{empty}

\begin{abstract}
  This is a tale of two linear programming decoders, namely channel coding
  linear programming decoding (CC-LPD) and compressed sensing linear
  programming decoding (CS-LPD). So far, they have evolved quite
  independently. The aim of the present paper is to show that there is a tight
  connection between, on the one hand, CS-LPD based on a zero-one measurement
  matrix over the reals and, on the other hand, CC-LPD of the binary linear
  code that is obtained by viewing this measurement matrix as a binary
  parity-check matrix. This connection allows one to translate performance
  guarantees from one setup to the other.
\end{abstract}

\section{Introduction}
\label{sec:introduction}

Recently there has been substantial interest in the theory of recovering
sparse approximations of signals that satisfy linear measurements. Compressed
(or compressive) sensing research (see, e.g., \cite{Candes:Tao:05:1,DonohoCS})
has developed conditions for measurement matrices under which (approximately)
sparse signals can be recovered by solving a linear programming relaxation of
the original NP-hard combinatorial problem. Interestingly, in one of the first
papers in this area (cf.~\cite{Candes:Tao:05:1}), Candes and Tao presented a
setup they called ``decoding by linear programming,'' henceforth called
\CSLPD, where the sparse signal corresponds to real-valued
noise that is added to a real-valued signal that is to be recovered in a
hypothetical communication problem.

At about the same time, in an independent line of research, Feldman,
Wainwright, and Karger considered the problem of decoding a binary linear code
that is used for data communication over a binary-input memoryless channel, a
problem that is also NP-hard in general. In~\cite{Feldman:03:1,
  Feldman:Wainwright:Karger:05:1}, they formulated this channel coding problem
as an integer linear program, along with presenting a linear programming
relaxation for it, henceforth called \CCLPD. Several theoretical results were
subsequently proven about the efficiency of \CCLPD, in particular for
low-density parity-check (LDPC) codes (e.g.~\cite{Koetter:Vontobel:03:1,
  Vontobel:Koetter:05:1:subm, Feldman:Malkin:Servedio:Stein:Wainwright:04:1,
  DDKW07}).

As we will see in the subsequent sections, \CSLPD\ and \CCLPD\ (and the setups
they are derived from) are \emph{formally} very similar, however, it is rather
unclear if there is a connection beyond this formal relationship. In fact
Candes and Tao in their original paper asked the following
question~\cite[Section VI.A]{Candes:Tao:05:1}: \emph{ ``\ldots In summary,
  there does not seem to be any explicit known connection with this line of
  work\footnote{Candes and Tao~\cite[Section VI.A]{Candes:Tao:05:1} refer here
    to~\cite{Feldman:03:1, Feldman:Wainwright:Karger:05:1}.} but it would
  perhaps be of future interest to explore if there is one.''}

In this paper we present such a connection between \CSLPD\ and \CCLPD. The
general form of our results is that if a given binary parity-check matrix is
``good'' for \CCLPD\ then the same matrix (considered over the reals) is a
``good'' measurement matrix for \CSLPD. The notion of a ``good'' parity-check
matrix depends on which channel we use (and a corresponding channel-dependent
quantity called pseudo-weight).

\begin{itemize}

\item Based on results for the binary symmetric channel (BSC), we show that if
  a parity-check matrix can correct any $k$ bit-flipping errors under \CCLPD,
  then the same matrix taken as a measurement matrix over the reals can be
  used to recover all $k$-sparse error signals under \CSLPD.

\item Based on results for binary-input output-symmetric channels with bounded
  log-likelihood ratios, we can extend the previous result to show that
  performance guarantees for \CCLPD\ for such channels can be translated into
  robust sparse-recovery guarantees in the $\ell_1 / \ell_1$ sense (see, e.g.,
  \cite{GIKS_RIP}) for \CSLPD.

\item Performance guarantees for \CCLPD\ for the binary-input AWGNC (additive
  white Gaussian noise channel) can be translated into robust sparse-recovery
  guarantees in the $\ell_2 / \ell_1$ sense for \CSLPD

\item Max-fractional weight performance guarantees for \CCLPD\ can be
  translated into robust sparse-recovery guarantees in the $\ell_{\infty} /
  \ell_1$ sense for \CSLPD.

\item Performance guarantees for \CCLPD\ for the BEC (binary erasure channel)
  can be translated into performance guarantees for the compressed sensing
  setup where the support of the error signal is known and the decoder tries
  to recover the sparse signal (i.e., tries to solve the linear equations) by
  back-substitution only.

\end{itemize}
All our results are also valid in a stronger, point-wise sense. For example,
for the BSC, if a parity-check matrix can recover a \emph{given set} of $k$
bit flips under \CCLPD, the same matrix will recover any sparse signal
supported on those $k$ coordinates under \CSLPD. In general, ``good''
performance of \CCLPD\ on a given error support will yield ``good'' \CSLPD\
recovery for sparse signals supported on the same support.
  
It should be noted that all our results are only one-way: we do not prove that
a ``good'' zero-one measurement matrix will always be a ``good'' parity-check
matrix for a binary code. This remains an interesting open problem.

The remainder of this paper is organized as follows. In
Section~\ref{sec:notation:1} we set up the notation that will be used. Then in
Sections~\ref{sec:cs:lpd:1} and~\ref{sec:cc:lpd:1} we will review the
compressed sensing and channel coding setups that we are interested in, along
with their respective linear programming relaxations. This review will be
presented in such a way that the close \emph{formal} relationship between the
two setups will stand out. Afterwards, in Section~\ref{sec:bridge:1} we will
show that for a zero-one matrix, once seen as a real-valued measurement
matrix, once seen as a binary parity-check matrix, this close relationship is
\emph{not only formal} but that in fact non-zero vectors in the real nullspace
of this matrix (i.e., vectors that are problematic vectors for \CSLPD) can be
mapped to non-zero vectors in the fundamental cone defined by that same matrix
(i.e., to vectors that are problematic vectors for \CCLPD). Based on this
observation one can, as will be shown in Section~\ref{sec:translation:1},
translate performance guarantees from one setup to the other. The paper
finishes with some conclusions in Section~\ref{sec:conclusions:1}.

\section{Basic Notation}
\label{sec:notation:1}

Let $\Z$, $\Zp$, $\Zpp$, $\R$, $\Rp$, $\Rpp$, and $\GF{2}$ be the ring of
integers, the set of non-negative integers, the set of positive integers, the
field of real numbers, the set of non-negative real numbers, the set of
positive real numbers, and the finite field of size $2$, respectively. Unless
noted otherwise, expressions, equalities, and inequalities will be over the
field $\R$. The absolute value of a real number $a$ will be denoted by
$|a|$. The size of a set $\setS$ will be denoted by $\card{\setS}$.

In this paper all vectors will be \emph{column} vectors. If $\vect{a}$ is some
vector with integer entries, then $\vect{a} \ (\mathrm{mod} \ 2)$ will denote
an equally long vector whose entries are reduced modulo $2$. If $\setS$ is a
subset of the set of coordinate indices of a vector $\va$ then $\va_{\setS}$
is the vector of length $\card{\setS}$ that contains only the coordinates of
$\va$ whose coordinate index appears in $\setS$. Moreover, if $\va$ is a real
vector then we define $|\va|$ to be the real vector $\va'$ of the same length
as $\va$ with entries $a'_i = |a_i|$ for all $i$. Finally, the inner product
$\langle \va, \vb \rangle$ of two equally long vectors $\va$ and $\vb$ is
defined to $\langle \va, \vb \rangle = \sum_i a_i b_i$.

We define $\supp(\va) \defeq \{ i \ | \ a_i \neq 0 \}$ to be the support set
of some vector $\va$. Moreover, we let $\setweightR{n}{k} \defeq \bigl\{ \va
\in \R^n \bigm| \card{\supp(\va)} \leq k \bigr\}$ and $\setweightGFtwo{n}{k}
\defeq \bigl\{ \va \in \GF{2}^n \bigm| \card{\supp(\va)} \leq k \bigr\}$ be
the set of vectors in $\R^n$ and $\GF{2}^n$, respectively, which have at most
$k$ non-zero components. If $k \ll n$ then vectors in these sets are called
$k$-sparse vectors.

For any real vector $\va$, we define $\zeronorm{\va}$ to be the $\ell_0$ norm
of $\va$, i.e., the number of non-zero components of $\va$. Note that
$\zeronorm{\va} = \wH(\va) = |\supp(\va)|$, where $\wH(\va)$ is the Hamming
weight of $\va$. Furthermore, $\onenorm{\va} \defeq \sum_i |a_i|$,
$\twonorm{\va} \defeq \sqrt{\sum_i |a_i|^2}$, $\infnorm{\va} \defeq \max_i
|a_i|$ will denote, respectively, the $\ell_1$, $\ell_2$, and $\ell_{\infty}$
norm of $\va$.

For a matrix $\matr{M}$ over $\R$ with $n$ columns we define its $\R$
nullspace to be the set $\nullspaceR(\matr{H}) \defeq \big\{ \va \in \R^n
\bigm| \matr{M} \cdot \va = \vzero \}$ and for a matrix $\matr{M}$ over
$\GF{2}$ with $n$ columns we define its $\GF{2}$ nullspace to be the set
$\nullspaceGFtwo(\matr{H}) \defeq \big\{ \va \in \GF{2}^n \bigm| \matr{M}
\cdot \va = \vzero \ \inGFtwo \}$.

Let $\matr{H} = (h_{j,i})_{j,i}$ be some matrix. We define the sets
$\setJ(\matr{H})$ and $\setI(\matr{H})$ to be, respectively, the set of row
and column indices of $\matr{H}$. Moreover, we will use the sets
$\setJ_i(\matr{H}) \defeq \{ j \in \setJ \ | \ h_{j,i} \neq 0 \}$ and
$\setI_j(\matr{H}) \defeq \{ i \in \setI \ | \ h_{j,i} \neq 0 \}$. In the
following, when no confusion can arise, we will sometimes omit the argument
$\matr{H}$ in the preceding expressions. For any set $\setS \subseteq \setI$,
we will denote its complement with respect to $\setI$ by $\setSc$, i.e.,
$\setSc \defeq \setI \setminus \setS$.

\section{Compressed Sensing \\ Linear Programming Decoding}
\label{sec:cs:lpd:1}

\subsection{The Setup}
\label{sec:cs:lpd:setup:1}

Let $\matrHCS$ be a real matrix of size $m \times n$, called the measurement
matrix, and let $\vs$ be a real vector of length $m$. In its simplest form,
the compressed sensing problem consists of finding the sparsest real vector
$\ve'$ of length $n$ that satisfies $\matrHCS \cdot \ve' = \vs$, namely

\optprog
{
\begin{alignat*}{2}
  \CSOPT:
  \quad
  &
  \text{minimize} \quad
  &&
    \zeronorm{\ve'} \\
  &
  \text{subject to } \quad
  &&
   \matrHCS \cdot \ve' = \vs.
\end{alignat*}
}{-0.8cm}

\noindent
Assuming that there exists a truly sparse signal $\ve$ that satisfies the
measurement $\matrHCS \cdot \ve = \vs$, \CSOPT\ yields, for suitable matrices
$\matrHCS$, an estimate $\hve$ that equals $\ve$.

This problem can also be interpreted~\cite{Candes:Tao:05:1} as part of the
decoding problem that appears in a coded data communicating setup where the
channel input alphabet is $\setXCS \defeq \R$, the channel output alphabet
is $\setYCS \defeq \R$, and the information symbols are encoded with the
help of a real-valued code $\codeCCS$ of length $n$ and dimension $\kappa
\defeq n - \rank_{\R}(\matrHCS)$ as follows.
\begin{itemize}

\item The code is $\codeCCS \defeq \bigl\{ \vx \in \R^n \bigm| \matrHCS \cdot
  \vx = \vzero \bigr\}$. Because of this, the measurement matrix $\matrHCS$ is
  sometimes also called an annihilator matrix.

\item A matrix $\matrGCS \in \R^{n \times \kappa}$ for which $\codeCCS =
  \bigl\{ \matrGCS \cdot \vu \bigm| \vu \in \R^{\kappa} \bigr\}$ holds, is
  called a generator matrix for the code $\codeCCS$. With the help of such a
  matrix, information vectors $\vu \in \R^{\kappa}$ are encoded into codewords
  $\vx \in \R^n$ according to $\vx = \matrGCS \cdot \vu$.

\item Let $\vy \in \setYCS^n$ be the received vector. We can write $\vy = \vx
  + \ve$ for a suitably defined vector $\ve \in \R^n$, which will be called
  the error vector. We assume that the channel is such that $\ve$ is sparse or
  approximately sparse.

\item The receiver first computes the syndrome vector $\vs$ according to $\vs
  \defeq \matrHCS \cdot \vy$. Note that
  \begin{align*}
    \vs
      &= \matrHCS \cdot (\vx + \ve)
       = \matrHCS \cdot \vx + \matrHCS \cdot \ve \\
      &= \matrHCS \cdot \ve.
  \end{align*}
  In a second step, the receiver solves \CSOPT\ to obtain an estimate $\hve$
  for $\ve$, which can be used to obtain the codeword estimate $\hvx = \vy -
  \hve$, which in turn can be used to obtain the information word estimate
  $\hvu$.
\end{itemize}

Because the complexity of solving \CSOPT\ is usually exponential in the
relevant parameters, one can try to formulate and solve a related optimization
problem with the aim that the related optimization problem yields very often
the same solution as \CSOPT, or at least very often a very good approximation
to the solution given by \CSOPT. In the context of \CSOPT, a popular approach
is to formulate and solve the following related optimization problem (which,
with the suitable introduction of auxiliary variables, can be turned into a
linear program):

\optprog
{
\begin{alignat*}{2}
  \CSLPD:
  \quad
  &
  \text{minimize} \quad
  &&
    \onenorm{\ve'} \\
  &
  \text{subject to } \quad
  &&
   \matrHCS \cdot \ve' = \vs.
\end{alignat*}
}{-0.8cm}

\subsection{Conditions for the Equivalence of \CSLPD\  and \CSOPT}

A central question of compressed sensing theory is under what conditions the
solution given by \CSLPD\ equals (or is very close to) the solution given by
\CSOPT.\footnote{It is important to note that we worry only about the solution
  given by \CSLPD\ being equal (or very close to) the solution given by
  \CSOPT, because even \CSOPT\ might fail to correctly estimate the error
  vector in the above communication setup when the error vector has too many
  large components.} Clearly, if $m \geq n$ and the matrix $\matrHCS$ has rank
$n$, there is only one feasible $\ve'$ and the two problems have the same
solution.

In this paper we typically focus on the linear sparsity regime, i.e., $k =
\Theta(n)$ and $m = \Theta(n)$, but our techniques are more generally
applicable. The question is for which measurement matrices (hopefully with a
small number of measurements $m$) the LP relaxation is tight, i.e., the
estimate given by $\CSLPD$ equals the estimate given by $\CSOPT$. One such
\emph{sufficient} condition is that a given measurement matrix is ``good'' if
it satisfies the restricted isometry property (RIP), i.e., does not distort
the $\ell_2$ length of all $k$-sparse vectors. If this is the case then it was
shown~\cite{Candes:Tao:05:1} that the LP relaxation will be tight for all
$k$-sparse vectors $\ve$ and further the recovery will be robust to
approximate sparsity. The RIP condition however is not a complete
characterization of ``good'' measurement matrices. We will use the nullspace
characterization (see, e.g.,
\cite{Xu:Hassibi:08:1:preprint,Stojnic:Xu:Hassibi:08:1}) instead, that is
necessary and sufficient.

\begin{definition}
  \label{def:nullspace:property:setS:1}

  Let $\setS \subseteq \setI(\matrHCS)$ and let $C \in \Rp$. We say that
  $\matrHCS$ has the nullspace property $\NSPR(\setS,C)$, and write $\matrHCS
  \in \NSPR(\setS,C)$, if
  \begin{align*}
    C \cdot \onenorm{\vnnuS}
      &\leq \onenorm{\vnnuSc}
       \ \ 
       \text{for all}
       \ \ 
       \vnnu \in \nullspaceR(\matrHCS).
  \end{align*}
  We say that $\matrHCS$ has the strict nullspace property $\SNSPR(\setS,C)$,
  and write $\matrHCS \in \SNSPR(\setS,C)$, if
  \begin{align*}
    C \cdot \onenorm{\vnnuS}
      &< \onenorm{\vnnuSc}
       \ \ 
       \text{for all}
       \ \ 
       \vnnu \in \nullspaceR(\matrHCS) \setminus \{ \vzero \}.\\[-0.75cm]
  \end{align*}
\end{definition}

\begin{definition}
  \label{def:nullspace:property:k:1}

  Let $k \in \Zp$ and let $C \in \Rp$. We say that $\matrHCS$ has the
  nullspace property $\NSPR(k,C)$, and write $\matrHCS \in \NSPR(k,C)$, if
  \begin{align*}
    \matrHCS
      &\in \NSPR(\setS,C)
       \ \ 
       \text{for all}
       \ \ 
       \text{$\setS \subseteq \setI(\matrHCS)$
             with $\card{\setS} \leq k$}.
  \end{align*}
  We say that $\matrHCS$ has the strict nullspace property $\SNSPR(k,C)$, and
  write $\matrHCS \in \SNSPR(k,C)$, if
  \begin{align*}
    \matrHCS
      &\in \SNSPR(\setS,C)
       \ \ 
       \text{for all}
       \ \ 
       \text{$\setS \subseteq \setI(\matrHCS)$
             with $\card{\setS} \leq k$}.\\[-0.75cm]
  \end{align*}
\end{definition}

As was shown independently by several authors (see \cite{Zhang_nullspace,
  Linial, FeuerCS, Stojnic:Xu:Hassibi:08:1} and references therein) the
nullspace condition in Definition~\ref{def:nullspace:property:k:1} is a
necessary and sufficient condition for a measurement matrix to be ``good'' for
$k$-sparse signals, i.e. that the estimate given by \CSLPD\ equals the
estimate given by \CSOPT\ for these matrices. The nullspace characterization
of ``good'' measurement matrices will be one of the keys to linking \CSLPD\
with \CCLPD. Observe that the requirement is that vectors in the nullspace of
$\matrHCS$ have their $\ell_1$ mass spread in substantially more than $k$
coordinates. The following theorem is adapted
from~\cite{Stojnic:Xu:Hassibi:08:1} (and references therein).

\begin{theorem}
  Let $\matrHCS$ be a measurement matrix. Further, assume that $\vs = \matrHCS
  \cdot \ve$ and that $\ve$ has at most $k$ nonzero elements, i.e.,
  $\zeronorm{\ve} \leq k$. Then the estimate $\hve$ produced by \CSLPD\ will
  equal the estimate $\hve$ produced by \CSOPT\ if $\matrHCS \in \SNSPR(k,
  C\!=\!1)$.
\end{theorem}

\noindent
\qquad \emph{Remark:} Actually, as discussed in~\cite{Stojnic:Xu:Hassibi:08:1}
and references therein, the condition $\matrHCS \in \SNSPR(k, C\!=\!1)$ is
also necessary, but we will not use this here.

The next performance metric (see, e.g., \cite{GIKS_RIP, CDD06}) for CS
involves recovering sparse approximations to signals that are not exactly
$k$-sparse.

\begin{definition}
  
  An $\ell_p/\ell_q$ approximation guarantee for \CSLPD\ means that the
  \CSLPD\ outputs an estimate $\hve$ that is within a factor $C_{p,q}(k)$ from
  the best $k$-sparse approximation for $\ve$, i.e.,
  \begin{align}
    \norm{\ve-\hve}{p}
      &\leq
        C_{p,q}(k)
        \cdot
        \min_{\ve' \in \setweightR{n}{k}}
          \norm{\ve-\ve'}{q},
            \label{eq:lp:lq;approximation:1}
  \end{align}
  where the left-hand side is measured in the $\ell_p$ norm and the right-hand
  side is measured in the $\ell_q$ norm.
\end{definition}

\noindent
Note that the minimizer of the right-hand side
of~\eqref{eq:lp:lq;approximation:1} (for any norm) is the vector $\ve' \in
\setweightR{n}{k}$ that has the $k$ largest (in magnitude) coordinates of
$\ve$, also called the best $k$-term approximation of
$\ve$~\cite{CDD06}. Therefore the right-hand side
of~\eqref{eq:lp:lq;approximation:1} equals $C_{p,q}(k) \cdot
\norm{\ve_{\overline{\setS^{*}}}}{q}$ where $\setS^{*}$ is the support set of
the $k$ largest (in magnitude) components of $\ve$. Also note that if $\ve$ is
exactly $k$-sparse the above condition suggests that $\hve = \ve$ since the
right hand-side of~\eqref{eq:lp:lq;approximation:1} vanishes, therefore it is
a strictly stronger statement than recovery of sparse signals. (Of course,
such a stronger approximation guarantee for $\hve$ is usually only obtained
under stronger assumptions on the measurement matrix.)

The nullspace condition is necessary and sufficient for $\ell_1/\ell_1$
approximation for any measurement matrix. This is shown in the next theorem
and proof which are adapted
from~\cite[Theorem~1]{Xu:Hassibi:08:1:preprint}. (Actually, we omit the
necessity part in the next theorem since it will not be needed in this paper.)

\begin{theorem}

  Let $\matrHCS$ be a measurement matrix and choose some constant $C >
  1$. Further, assume that $\vs = \matrHCS \cdot \ve$. Then for any set $\setS
  \subseteq \setI$ with $\card{\setS} \leq k$ the solution $\hve$ produced by
  \CSLPD\ will satisfy
  \begin{align*}
    \onenorm{\ve - \hve}
      &\leq 2
            \cdot
            \frac{C+1}{C-1}
            \cdot
            \onenorm{\veSc}
  \end{align*}
  if $\matrHCS \in \NSPR(k,C)$.
\end{theorem}

\begin{proof}
  Suppose that $\matrHCS$ has the claimed nullspace property.
  Since $\matrHCS \cdot \ve = \vs$ and $\matrHCS \cdot \hve = \vs$, it easily
  follows that $\vnnu \defeq \ve - \hve$ is in the nullspace of
  $\matrHCS$. So,
  \begin{align}
    \onenorm{\veS}
    +
    \onenorm{\veSc}
      &= \onenorm{\ve} \nonumber \\
      &\overset{\text{(a)}}{\geq}
         \onenorm{\hve} \nonumber \\
      &= \onenorm{\ve + \vnnu} \nonumber \\
      &= \onenorm{\veS + \vnnuS}
         + 
         \onenorm{\veSc + \vnnuSc} \nonumber \\
      &\overset{\text{(b)}}{\geq}
         \onenorm{\veS}
         -
         \onenorm{\vnnuS}
         +
         \onenorm{\vnnuSc}
         -
         \onenorm{\veSc} \nonumber \\
      &\overset{\text{(c)}}{\geq}
         \onenorm{\veS}
         +
         \frac{C-1}{C+1}
           \cdot
           \onenorm{\vnnu}
         -
         \onenorm{\veSc},
           \label{eq:ell1:ell1:guarantee:1}
  \end{align}
  where step~$\text{(a)}$ follows from the fact that the solution to \CSLPD\
  satisfies $\onenorm{\hve} \leq \onenorm{\ve}$, where step~$\text{(b)}$
  follows from applying the triangle inequality for the $\ell_1$ norm twice,
  and where step~$\text{(c)}$ follows from
  \begin{align*}
    -
    \onenorm{\vnnuS}
    +
    \onenorm{\vnnuSc}
      &\overset{\text{(d)}}{\geq}
         \frac{C-1}{C+1}
         \cdot
         \onenorm{\vnnu}.
  \end{align*}
  Here, step~$\text{(d)}$ is a consequence of
  \begin{align*}
    (C{+}1)
      \, \cdot \, &
      \big(
        -
        \onenorm{\vnnuS}
        +
        \onenorm{\vnnuSc}
      \big) \\
      &= -
         C
           \cdot
           \onenorm{\vnnuS}
         -
         \onenorm{\vnnuS}
         +
         C
           \cdot
           \onenorm{\vnnuSc}
         +
         \onenorm{\vnnuSc} \\
      &\overset{\text{(e)}}{\geq}
         -
         \onenorm{\vnnuSc}
         -
         \onenorm{\vnnuS}
         +
         C
           \cdot
           \onenorm{\vnnuSc}
         +
         C
           \cdot
           \onenorm{\vnnuS} \\
      &= (C{-}1)
           \cdot
           \onenorm{\vnnuS}
         +
         (C{-}1)
           \cdot
           \onenorm{\vnnuSc} \\
      &= (C{-}1)
           \cdot
           \onenorm{\vnnu},
  \end{align*}
  where step~$\text{(e)}$ follows from applying twice the fact that $\vnnu \in
  \nullspaceR(\matrHCS)$ and the assumption that $\matrHCS \in
  \NSPR(k,C)$. Subtracting the term $\onenorm{\veS}$ on both sides
  of~\eqref{eq:ell1:ell1:guarantee:1}, and solving for $\onenorm{\vnnu} =
  \onenorm{\ve - \hve}$ yields the promised result.
\end{proof}

\section{Channel Coding \\ Linear Programming Decoding}
\label{sec:cc:lpd:1}

\subsection{The Setup}
\label{sec:cc:lpd:setup:1}

We consider coded data transmission over a memoryless channel with input
alphabet $\setXCC \defeq \{ 0, 1 \}$, output alphabet $\setYCC$, and channel
law $P_{Y | X}(y | x)$ with the help of a binary linear code $\codeCCC$ of
length $n$ and dimension $\kappa$ with $n \geq \kappa$. In the following, we
will identify $\setXCC$ with $\GF{2}$.
\begin{itemize}

\item Let $\matrGCC \in \GF{2}^{n \times \kappa}$ be a generator matrix for
  $\codeCCC$. Consequently, $\matrGCC$ has rank $\kappa$ over $\GF{2}$, and
  information vectors $\vu \in \GF{2}^{\kappa}$ are encoded into codewords
  $\vx \in \GF{2}^n$ according to $\vx = \matrGCC \cdot \vu \ \inGFtwo$,
  i.e.,. $\codeCCC = \bigl\{ \matrGCC \cdot \vu \ \inGFtwo \bigm| \vu \in
  \GF{2}^{\kappa} \bigr\}$.\footnote{We remind the reader that throughout this
    paper we are using \emph{column} vectors, which is in contrast to the
    coding theory habit to use \emph{row} vectors.}

\item Let $\matrHCC \in \GF{2}^{m \times n}$ be a parity-check matrix for
  $\codeCCC$. Consequently, $\matrHCC$ has rank $n-\kappa \leq m$ over
  $\GF{2}$, and any $\vx \in \GF{2}^n$ satisfies $\matrHCC \cdot \vx = \vzero \
  \inGFtwo$ if and only if $\vx \in \codeCCC$, i.e., $\codeCCC = \bigl\{ \vx
  \in \GF{2}^n \bigm| \matrHCC \cdot \vx = \vzero \ \inGFtwo \bigr\}$.

\item Let $\vy \in \setYCC^n$ be the received vector and define for each $i
  \in \setI(\matrHCC)$ the log-likelihood ratio $\lambda_i \defeq
  \lambda_i(y_i) \defeq \log \bigl( \frac{P_{Y | X}(y_i|0)}{P_{Y | X}(y_i |
    1)} \bigr)$.

\item On the side, let us remark that if $\setYCC$ is binary then $\setYCC$
  can be identified with $\GF{2}$ and we can write $\vy = \vx + \ve \
  \inGFtwo$ for a suitably defined vector $\ve \in \GF{2}^n$, which will be
  called the error vector. Moreover, we can define the syndrome vector $\vs
  \defeq \matrHCC \cdot \vy \ \inGFtwo$. Note that
  \begin{align*}
    \vs
      &= \matrHCC \cdot (\vx + \ve)
       = \matrHCC \cdot \vx + \matrHCC \cdot \ve \\
      &= \matrHCC \cdot \ve
         \quad \inGFtwo.
  \end{align*}
  However, in the following we will only use the log-likelihood ratio vector
  $\vlambda$ (that can be defined for any alphabet $\setYCC$), and not the
  binary syndrome vector $\vs$.
\end{itemize}
Upon observing $\vect{Y} = \vect{y}$, the maximum-likelihood decoding (MLD)
rule decides for $\hat \vx(\vy) = \arg \max_{\vx' \in \codeCCC} P_{\vY |
  \vX}(\vy | \vx')$ where $P_{\vY |\vX}(\vy | \vx') = \prod_{i \in \set{I}}
P_{Y | X}(y_i | x'_i)$.\footnote{Actually, slightly more precise would be to
  call this decision rule ``block-wise maximum-likelihood decoding.''}
Formally:

\optprog
{
\begin{alignat*}{2}
  \CCMLDone:
  \quad
  &
  \text{maximize} \quad
  &&
    P_{\vY |\vX}(\vy | \vx') \\
  &
  \text{subject to } \quad
  &&
   \vx' \in \codeCCC.
\end{alignat*}
}{-0.8cm}

\noindent It is clear that instead of $P_{\vY |\vX}(\vy | \vx')$ we can also
maximize $\log P_{\vY |\vX}(\vy | \vx') = \sum_{i \in \set{I}} \log P_{Y |
  X}(y_i | x'_i)$. Noting that $\log P_{Y | X}(y_i | x'_i) = - \lambda_i x'_i
+ \log P_{Y | X}(y_i | 0)$ for $x'_i \in \{ 0, 1 \}$, $\CCMLDone$ can then be
rewritten to read

\optprog
{
\begin{alignat*}{2}
  \CCMLDtwo:
  \quad
  &
  \text{minimize} \quad
  &&
    \langle \vlambda, \vx' \rangle \\
  &
  \text{subject to } \quad
  &&
   \vx' \in \codeCCC.
\end{alignat*}
}{-0.8cm}

\noindent Because the cost function is linear, and a linear function attains
its minimum at the extremal points of a convex set, this is essentially
equivalent to

\optprog
{
\begin{alignat*}{2}
  \CCMLDthree:
  \quad
  &
  \text{minimize} \quad
  &&
    \langle \vlambda, \vx' \rangle \\
  &
  \text{subject to } \quad
  &&
   \vx' \in \convhull(\codeCCC).
\end{alignat*}
}{-0.8cm}

\noindent Although this is a linear program, it can usually not be solved
efficiently because its description complexity is typically exponential in the
block length of the code.\footnote{Examples of code families that have
  sub-exponential description complexities in the block length are
  convolutional codes (with fixed state-space size), cycle codes, and tree
  codes. However, these classes of codes are not good enough for achieving
  performance close to capacity even under ML decoding. (For more on this
  topic, see for example~\cite{Kashyap:08:1}.)}

However, one might try to solve a relaxation of \CCMLDthree. Namely, as
proposed by Feldman, Wainwright, and Karger~\cite{Feldman:03:1,
  Feldman:Wainwright:Karger:05:1}, we can try to solve the optimization
problem

\optprog
{
\begin{alignat*}{2}
  \CCLPD:
  \quad
  &
  \text{minimize} \quad
  &&
    \langle \vlambda, \vx' \rangle \\
  &
  \text{subject to } \quad
  &&
   \vx' \in \fp{P}(\matrHCC),
\end{alignat*}
}{-0.8cm}

\noindent
where the relaxed set $\fp{P}(\matrHCC) \supseteq \convhull(C)$ is given in
the next definition.

\begin{definition}
  For every $j \in \setJ(\matrHCC)$, let $\vect{h}^\tr_j$ be the $j$-th row of
  $\matrHCC$ and let $\codeCCCdown{j} \defeq \bigl\{ \vx \in \GF{2}^n \bigm|
  \langle \vect{h}_j, \vx \rangle = 0 \text{ (mod $2$)} \bigr\}$. Then, the
  fundamental polytope $\fp{P} \defeq \fp{P}(\matrHCC)$ of $\matrHCC$ is
  defined to be the set
  \begin{align*}
    \fp{P}
      &\defeq
    \fp{P}(\matrHCC)
       = \bigcap_{j \in \setJ} \convhull(\codeCCCdown{j}).
  \end{align*}
  Vectors in $\fp{P}(\matrHCC)$ will be called pseudo-codewords.
\end{definition}

In order to motivate this relaxation, note that the code $\code{C}$ can be
written as
\begin{align*}
  \codeCCC
    &= \codeCCCdown{1} \cap \cdots \cap \codeCCCdown{m},
\end{align*}
and so
\begin{align*}
  \convhull(\codeCCC)
    &= \convhull
         (
           \codeCCCdown{1} \cap \cdots \cap \codeCCCdown{m}
         ) \\
    &\subseteq
       \convhull
         (
           \codeCCCdown{1}) \cap \cdots \cap \convhull(\codeCCCdown{m}
         ) \\
    &= \fp{P}(\matrHCC).
\end{align*}
It can be verified~\cite{Feldman:03:1, Feldman:Wainwright:Karger:05:1} that
this relaxation possesses the important property that all the vertices of
$\convhull(\codeCCC)$ are also vertices of $\fp{P}(\matrHCC)$. Let us
emphasize that different parity-check matrices for the same code usually lead
to different fundamental polytopes and therefore to different \CCLPD{}s.

Similarly to the compressed sensing setup, we want to understand when we can
guarantee that the codeword estimate given by $\CCLPD$ equals the codeword
estimate given by $\CCMLD$. It is important to note, as we did in the
compressed sensing setup, that we worry mostly about the solution given by
\CCLPD\ being equal to the solution given by \CCMLD, because even \CCMLD\
might fail to correctly identify the codeword that was sent when the error
vector is beyond the error correction capability of the code. Therefore, the
performance of \CCMLD\ is a natural upper bound on the performance of \CCLPD,
and a way to assess \CCLPD\ is to study the gap to \CCMLD, e.g., by comparing
the performance guarantees for \CCLPD\ that are discussed here with known
performance guarantees for \CCMLD.

When characterizing the \CCLPD\ performance of binary linear codes over
binary-input output-symmetric channels~\cite{Richardson:Urbanke:08:1} we can
without loss of generality assume that the all-zero codeword was
transmitted. With this, the success probability of \CCLPD\ is the probability
that the all-zero codeword yields the lowest cost function value compared to
all non-zero vectors in the fundamental polytope. Because the cost function is
linear, this is equivalent to the statement that the success probability of
\CCLPD\ equals the probability that the all-zero codeword yields the lowest
cost function value compared to all non-zero vectors in the conic hull of the
fundamental polytope. This conic hull is called the fundamental cone $\fc{K}
\defeq \fp{K}(\matrHCC)$ and it can be written as
\begin{align*}
  \fc{K}
    &\defeq
  \fc{K}(\matrHCC)
     = \conichull
         \big(
           \fp{P}(\matrHCC)
         \big)
     = \bigcap_{j \in \setJ} \conichull(\codeCCCdown{j}).
\end{align*}
The fundamental cone can be characterized by the inequalities listed in the
following lemma~\cite{Feldman:03:1, Feldman:Wainwright:Karger:05:1,
  Koetter:Vontobel:03:1, Vontobel:Koetter:05:1:subm}. (Similar inequalities
can be given for the fundamental polytope but we will not need them here.)

\begin{lemma}
  \label{lemma:fundamental:cone:1}

  The fundamental cone $\fc{K} \defeq \fc{K}(\matrHCC)$ of $\matrHCC$ is the
  set of all vectors $\vomega \in \R^n$ that satisfy
  \begin{alignat}{2}
    \omega_i
      &\geq 0 
      \ \ \ 
      &&\text{(for all $i \in \setI$)} \; , 
          \label{eq:fund:cone:def:1} \\
    \omega_i
      &\leq
          \sum_{i' \in \setI_j \setminus i} \!\!
            \omega_{i'}
      \ \ \ 
      &&\text{(for all $j \in \setJ$, \ 
               for all $i \in \setI_j$)} \; .
          \label{eq:fund:cone:def:2}
  \end{alignat}
  A vector $\vomega \in \fc{K}$ is called a pseudo-codeword. If such a vector
  lies on an edge of $\fc{K}$, it is called a minimal
  pseudo-codeword. Moreover, if $\vomega \in \fc{K} \cap \Z^n$ and $\vomega \
  (\mathrm{mod} \ 2) \in \code{C}$, then $\vomega$ is called an unscaled
  pseudo-codeword. (For a motivation of these definitions,
  see~\cite{Vontobel:Koetter:05:1:subm, Koetter:Li:Vontobel:Walker:07:1}).
\end{lemma}

Note that in the following, not only vectors in the fundamental polytope, but
also vectors in the fundamental cone will be called
pseudo-codewords. Moreover, if $\matrHCS$ is a zero-one measurement matrix,
i.e., a measurement matrix where all entries are in $\{ 0, 1 \}$, then we will
consider $\matrHCS$ to represent also the parity-check matrix of some linear
code over $\GF{2}$. Consequently, its fundamental polytope will be denoted by
$\fp{P}(\matrHCS)$ and its fundamental cone by $\fc{K}(\matrHCS)$.

\subsection{Conditions for the Equivalence of \CCLPD\  and \CCMLD}

The following lemma states when \CCLPD\ succeeds for the BSC.

\begin{lemma}
  \label{lemma:bsc:strict:balancedness:1}

  Let $\matrHCC$ be the parity-check matrix of some code $\codeCCC$ and let
  $\setS \subseteq \setI(\matrHCC)$ be the set of coordinate indices that are
  flipped by the BSC. If $\matrHCC$ is such that
  \begin{align}
    \onenorm{\vomega_{\setS}}
        &< \onenorm{\vomega_{\setSc}}
             \label{eq:bsc:strict:balancedness:1}
  \end{align} 
  for all $\vomega \in \fc{K}(\matrHCC) \setminus \{ \vzero \}$ then the
  \CCLPD\ decision equals the codeword that was sent.
\end{lemma}

\noindent
\qquad \emph{Remark:} The above condition is also necessary, however, we will
not use this fact in the following.

\begin{proof}
  Without loss of generality, we can assume that the all-zero codeword was
  transmitted. Let $+L > 0$ be the log-likelihood ratio associated to a
  received $0$, and let $-L < 0$ be the log-likelihood ratio associated to a
  received $1$. Therefore, $\lambda_i = +L$ if $i \in \setSc$ and $\lambda_i =
  -L$ if $i \in \setS$. Then it follows from the assumptions in the lemma
  statement that for any $\vomega \in \fc{K}(\matrHCC) \setminus \{ \vzero \}$
  \begin{align*}
     \langle \vlambda, \vomega \rangle
       &= \sum_{i \in \setSc}
            (+L) \cdot \omega_i
          +
          \sum_{i \in \setS}
            (-L) \cdot \omega_i \\
       &\overset{\text{(a)}}{=}
          L
            \cdot
            \onenorm{\vomega_{\setSc}}
          -
          L
            \cdot
            \onenorm{\vomega_{\setS}}
        \overset{\text{(b)}}{>}
          0
        = \langle \vlambda, \vzero \rangle,
  \end{align*}
  where the equality follows from the fact that $|\omega_i| = \omega_i$ for
  all $i \in \setI(\matrHCC)$, and where the inequality in step~$\text{(b)}$
  follows from~\eqref{eq:bsc:strict:balancedness:1}. Therefore, under \CCLPD\
  the all-zero codeword has the lowest cost function value compared to all the
  non-zero pseudo-codewords in the fundamental cone, and therefore also
  compared to all the non-zero pseudo-codewords in the fundamental polytope.
\end{proof}

Note that the inequality in~\eqref{eq:bsc:strict:balancedness:1} is
\emph{identical} to the inequality that appears in the definition of the
strict nullspace property for $C = 1$\ (!) This observation makes one wonder
if there is a connection between \CSLPD\ and \CCLPD, in particular for
measurement matrices that contain only zeros and ones. Of course, in order to
establish such a connection we first need to understand how points in the
nullspace of the measurement matrix $\matrHCS$ can be associated with points
in the fundamental polytope of the parity-check matrix $\matrHCS$ (now seen as
a parity-check matrix for a code over $\GF{2}$). Such an association will be
exhibited in Section~\ref{sec:bridge:1}. However, before turning to that
section, we will first discuss pseudo-weights, which are a popular way of
characterizing the importance of the different pseudo-codewords in the
fundamental cone and for establishing performance guarantees for \CCLPD.

\subsection{Definition of Pseudo-Weights}
\label{sec:pseudo:weight:definitions:1}

Note that the fundamental polytope and cone are only a function of
the parity-check matrix of the code and \emph{not} of the channel. The
influence of the channel is reflected in the pseudo-weight of the
pseudo-codewords, so every channel has its pseudo-weight definition.
Therefore, every communication channel comes with the right measure of
distance that determines how often a fractional vertex is incorrectly chosen
in \CCLPD.

\begin{definition}[\cite{Wiberg:96, Forney:Koetter:Kschischang:Reznik:01:1,
    Feldman:03:1, Feldman:Wainwright:Karger:05:1, Koetter:Vontobel:03:1,
    Vontobel:Koetter:05:1:subm}]
  \label{def:pseudo:weights:1}

  Let $\vomega$ be a non\-zero vector in $\Rp^n$ with $\vomega = (\omega_1,$
  $\ldots, \omega_n)$.
  \begin{itemize}
  
  \item The AWGNC (more precisely, binary-input AWGNC) pseudo-weight of
    $\vomega$ is defined to be
    \begin{align*}
      \wpsAWGNC(\vomega)
        &\defeq \frac{\onenorm{\vomega}^2}
                     {\twonorm{\vomega}^2}.
    \end{align*}

  \item In order to define the BSC pseudo-weight $\wpsBSC(\vomega)$, we let
    $\vomega'$ be the vector of length $n$ with the same components as
    $\vomega$ but in non-increasing order. Now let
    \begin{align*}
      f(\xi)
        &\defeq \omega'_i \quad (i-1 < \xi \leq i,\ 0 < \xi \leq n), \\
      F(\xi)
        &\defeq \int_{0}^{\xi} f(\xi') \dint{\xi'}, \\
      e
        &\defeq
           F^{-1} \left( \frac{F(n)}{2} \right)
         = F^{-1} \left( \frac{\onenorm{\vomega}}{2} \right).
    \end{align*}
    Then the BSC pseudo-weight $\wpsBSC(\vomega)$ of $\vomega$ is defined to
    be $\wpsBSC(\vomega) \defeq 2e$.

  \item The BEC pseudo-weight of $\vomega$ is defined to be
    \begin{align*}
      \wpsBEC(\vomega)
        &= \big| \supp(\vomega) \big|.
    \end{align*}

  \item The max-fractional weight of $\vomega$ is defined to be
    \begin{align*}
      \wmaxfr(\vomega)
        &\defeq
           \frac{\onenorm{\vomega}}
                {\infnorm{\vomega}}.
    \end{align*}

  \end{itemize}
  For $\vomega = \vzero$ we define all of the above pseudo-weights and the
  max-fractional weight to be zero.
\end{definition}

A detailed discussion of the motivation and significance of these definitions
can be found in~\cite{Vontobel:Koetter:05:1:subm}. For a parity-check matrix
$\matrHCC$ we define the minimum AWGNC pseudo-weight $\wpsAWGNCmin(\matrHCC)$
to be
\begin{align*}
  \wpsAWGNCmin(\matrHCC)
    &\defeq
       \min_{\vomega \in \fp{P}(\matrHCC) \setminus \{ \vzero \}}
         \wpsAWGNC(\vomega) \\
    &= \min_{\vomega \in \fc{K}(\matrHCC) \setminus \{ \vzero \}}
         \wpsAWGNC(\vomega).
\end{align*}
The minimum BSC pseudo-weight $\wpsBSCmin(\matrHCC)$, the minimum BEC
pseudo-weight $\wpsBECmin(\matrHCC)$, and the minimum max-fractional weight
$\wmaxfrmin(\matrHCC)$ of $\matrHCC$ are defined analogously. Note that
although $\wmaxfrmin(\matrHCC)$ yields weaker performance guarantees than the
other quantities~\cite{Vontobel:Koetter:05:1:subm}, it has the advantage of
being efficiently computable~\cite{Feldman:03:1,
  Feldman:Wainwright:Karger:05:1}.

There are other possible definitions of a BSC pseudo-weight. For example, the
BSC pseudo-weight of $\vomega$ can also be taken to be
\begin{align*}
  \wpsBSCmod(\vomega)
    &\defeq
       \begin{cases}
         2e     & \text{if $\onenorm{\vomega'_{\{ 1, \ldots, e \}}}
                    = \onenorm{\vomega'_{\{ e+1, \ldots, n \}}}$} \\
         2e - 1 & \text{if $\onenorm{\vomega'_{\{ 1, \ldots, e \}}}
                    > \onenorm{\vomega'_{\{ e+1, \ldots, n \}}}$}
       \end{cases},
\end{align*}
where $\vomega'$ is defined as in Definition~\ref{def:pseudo:weights:1} and
where $e$ is the smallest integer such that $\onenorm{\vomega'_{\{ 1, \ldots,
    e \}}} \geq \onenorm{\vomega'_{\{ e+1, \ldots, n \}}}$. This definition of
the BSC pseudo-weight was e.g.\ used in~\cite{Kelley:Sridhara:07:1}. (Note
that in~\cite{Forney:Koetter:Kschischang:Reznik:01:1} the quantity
$\wpsBSCmod(\vomega)$ was introduced as ``BSC effective weight''.)

Of course, the values $\wpsBSC(\vomega)$ and $\wpsBSCmod(\vomega)$ are tightly
connected. Namely, if $\wpsBSCmod(\vomega)$ is an even integer then
$\wpsBSCmod(\vomega) = \wpsBSC(\vomega)$, and if $\wpsBSCmod(\vomega)$ is an
odd integer then $\wpsBSCmod(\vomega) - 1 < \wpsBSC(\vomega) <
\wpsBSCmod(\vomega) + 1$.

The following lemma establishes a connection between BSC pseudo-weights and
the condition that appears in Lemma~\ref{lemma:bsc:strict:balancedness:1}.

\begin{lemma}
  \label{lemma:meaning:of:bsc:pseudo:weight:1}

  Let $\matrHCC$ be the parity-check matrix of some code $\codeCCC$ and let
  $\vomega$ be some arbitrary non-zero pseudo-codeword of $\matrHCC$, i.e.,
  $\vomega \in \fc{K}(\matrHCC) \setminus \{ \vzero \}$. Then for all sets
  $\setS \subseteq \setI$ with $\card{\setS} < \frac{1}{2} \cdot
  \wpsBSC(\vomega)$, or with $\card{\setS} < \frac{1}{2} \cdot
  \wpsBSCmod(\vomega)$, it holds that
  \begin{align*}
    \onenorm{\vomega_{\setS}}
      &< \onenorm{\vomega_{\setSc}}.
  \end{align*}
\end{lemma}

\begin{proof}
  First, consider the statement under for the assumption $\card{\setS} <
  \frac{1}{2} \cdot \wpsBSC(\vomega)$. The proof is by contradiction. So,
  assume that $\onenorm{\vomega_{\setS}} \geq \onenorm{\vomega_{\setSc}}$
  holds. This statement is clearly equivalent to the statement that $2 \cdot
  \onenorm{\vomega_{\setS}} \geq \onenorm{\vomega_{\setS}} +
  \onenorm{\vomega_{\setSc}} = \onenorm{\vomega}$, which is equivalent to the
  statement that $\onenorm{\vomega_{\setS}} \geq \frac{1}{2} \cdot
  \onenorm{\vomega}$. In terms of the notation in
  Definition~\ref{def:pseudo:weights:1}, this means that
  \begin{align*}
    \wpsBSC(\vomega)
      &= 2 \cdot F^{-1}\left( \frac{\onenorm{\vomega}}{2} \right)
       \overset{\text{(a)}}{\leq}
         2 \cdot F^{-1}(\onenorm{\vomega_{\setS}}) \\
      &\overset{\text{(b)}}{\leq}
         2
         \cdot
         \frac{\onenorm{\vomega_{\setS}}}{\infnorm{\vomega}}
       \leq
         2
         \cdot
         \frac{\card{\setS} \cdot \infnorm{\vomega}}{\infnorm{\vomega}}
       = 2
         \cdot
         \card{\setS},
  \end{align*}
  where at step~$\text{(a)}$ we have used the fact that $F^{-1}$ is a
  (strictly) non-decreasing function and where at step~$\text{(b)}$ we have
  used the fact that the slope of $F^{-1}$ (over the domain where $F^{-1}$ is
  defined) is at least $1 / \infnorm{\vomega}$. This, however, is a
  contradiction to the assumption that $\card{\setS} < \frac{1}{2} \cdot
  \wpsBSC(\vomega)$.

  Secondly, consider the statement under for the assumption $\card{\setS} <
  \frac{1}{2} \cdot \wpsBSCmod(\vomega)$. The proof is by contradiction. So,
  assume that the $\onenorm{\vomega_{\setS}} \geq \onenorm{\vomega_{\setSc}}$
  holds. With this, and the above definition of $\vomega'$ based on $\vomega$,
  $\onenorm{\vomega'_{\{ 1, \ldots, \card{\setS} \}}} \geq
  \onenorm{\vomega_{\setS}} \geq \onenorm{\vomega_{\setSc}} \geq
  \onenorm{\vomega'_{\{ \card{\setS}+1, \ldots, n \}}}$. If
  $\wpsBSCmod(\vomega)$ is an even integer then this line of inequalities
  shows that $\card{\setS} \geq \frac{1}{2} \cdot \wpsBSCmod(\vomega)$, which
  is a contradiction to the assumption that $\card{\setS} < \frac{1}{2} \cdot
  \wpsBSCmod(\vomega)$. If $\wpsBSCmod(\vomega)$ is an odd integer then this
  line of inequalities shows that $\card{\setS} \geq \frac{1}{2} \cdot
  \bigl(\wpsBSCmod(\vomega) + 1 \bigr) > \frac{1}{2} \wpsBSCmod(\vomega)$,
  which again is a contradiction to the assumption that $\card{\setS} <
  \frac{1}{2} \cdot \wpsBSCmod(\vomega)$.
\end{proof}

\section{Establishing a Bridge Between \\
               \CSLPD\  and \CCLPD}
\label{sec:bridge:1}

We are now ready to establish a bridge between
\CSLPD\  and \CCLPD. Our main tool is a simple lemma that was already
established in~\cite{Smarandache:Vontobel:09:2} but for a different purpose.

\begin{lemma}
  \label{lemma:equation:nullspace:to:fc:1}
 
  Let $\matrHCS$ be a measurement matrix that contains only zeros and
  ones. Then
  \begin{align*}
    \vnu \in \nullspaceR(\matrHCS)
    \ \ \ \Rightarrow \ \ \ 
    |\vnu| \in \fc{K}(\matrHCS).
  \end{align*}
\end{lemma}

\noindent
\qquad \emph{Remark:} Note that $\supp(\vnu) = \supp(|\vnu|)$.

\begin{proof}
  Let $\vomega \defeq |\vnu|$. In order to show that such a vector $\vomega$
  is indeed in the fundamental cone of $\matrHCS$, we need to
  verify~\eqref{eq:fund:cone:def:1} and~\eqref{eq:fund:cone:def:2}. The way
  $\vomega$ is defined, it is clear that it
  satisfies~\eqref{eq:fund:cone:def:1}. Therefore, let us focus on the proof
  that $\vomega$ satisfies~\eqref{eq:fund:cone:def:2}. Namely, from $\vnu \in
  \nullspaceR(\matrHCS)$ it follows that for all $j \in \setJ$, $\sum_{i \in
    \setI} h_{j,i} \nu_i = 0$, i.e., for all $j \in \setJ$, $\sum_{i \in
    \setI_j} \nu_i = 0$. This implies
  \begin{align*}
    \omega_i
      &= |\nu_i|
       = \left|
           \,\, - \!\!
           \sum_{i' \in \setI_j \setminus i}
             \nu_{i'}
         \right|
       \leq
         \sum_{i' \in \setI_j \setminus i}
            |\nu_{i'}|
       = \sum_{i' \in \setI_j \setminus i}
            \omega_{i'}
  \end{align*}
  for all $j \in \setJ$ and all $i \in \setI_j$, showing that $\vomega$
  indeed satisfies~\eqref{eq:fund:cone:def:2}.
\end{proof}

This lemma is fundamentally one-way: it says that with every point in the real
nullspace of the measurement matrix $\matrHCS$ we can associate a point in the
fundamental cone of $\matrHCS$, but not necessarily vice-versa. Therefore a
problematic point for the real nullspace of $\matrHCS$ will translate to a
problematic point in the fundamental cone of $\matrHCS$ and hence to bad
performance of \CCLPD. Similarly, a ``good'' parity-check matrix $\matrHCS$
must have no low pseudo-weight points in the fundamental cone, which means
that there are no problematic points in the real nullspace of $\matrHCS$.
Therefore ``positive'' results for channel coding will translate into
``positive'' results for compressed sensing, and ``negative'' results for
compressed sensing will translate into ``negative'' results for channel
coding.

Further, the lemma preserves the support of a given point $\vnu$. That means
that if there are no low pseudo-weight points in the fundamental cone of
$\matrHCS$ with a given support, there are no problematic points in the real
nullspace of $\matrHCS$ with the same support, which allows point-wise
versions of all our results.

\section{Translation of Performance Guarantees}
\label{sec:translation:1}

In this section we use the bridge between \CSLPD\ and \CCLPD\ that was
established in the previous section to translate ``positive'' results about
\CCLPD\ to ``positive'' results about \CSLPD.

\subsection{The Role of the BSC Pseudo-Weight for \CSLPD}

\begin{lemma}
  \label{lemma:from:bsc:to:cs:lpd:1}

  Let $\matrHCS \in \{ 0, 1 \}^{m \times n}$ be a CS measurement matrix and
  let $k$ be a non-negative integer. Then
  \begin{align*}
    \wpsBSCmin(\matrHCS) > 2k
    \ \ \ \Rightarrow \ \ \ 
    \matrHCS \in \SNSPR(k, C\!=\!1).
  \end{align*}
\end{lemma}

\begin{proof}
  Fix some $\vnu \in \nullspaceR(\matrHCS) \setminus \{ \vzero \}$. By
  Lemma~\ref{lemma:equation:nullspace:to:fc:1} we know that $|\vnu|$ is a
  pseudo-codeword of $\matrHCS$, and by the assumption $\wpsBSCmin(\matrHCS) >
  2k$ we know that $\wpsBSC(|\vnu|) > 2k$. Then, using
  Lemma~\ref{lemma:meaning:of:bsc:pseudo:weight:1}, we conclude that for all
  sets $\setS \subseteq \setI$ with $\card{\setS} \leq k$, we must have
  $\onenorm{\vnuS} = \onenorm{\,|\vnuS|\,} < \onenorm{\,|\vnuSc|\,} =
  \onenorm{\vnuSc}$. Because $\vnu$ was arbitrary, the claim $\matrHCS \in
  \SNSPR(k,C\!=\!1)$ clearly follows.
\end{proof}

Recent results on the performance analysis of \CCLPD\ showed that parity-check
matrices constructed from expander graphs can correct a constant fraction (of
the block length $n$) of worst
case~\cite{Feldman:Malkin:Servedio:Stein:Wainwright:07:1} and
random~\cite{DDKW07,ADS:Improved_LP} errors. (These types of results are
analogous to the so-called strong and weak bounds for compressed sensing,
respectively.)

These worst case error performance guarantees implicitly show that the BSC
pseudo-weight of all pseudo-codewords of a binary linear code defined by a
Tanner with sufficient expansion (strictly larger than $3/4$) must grow
linearly in $n$. (A conclusion in a similar direction can be drawn for the
random error setup.) We can therefore use our results to obtain new
performance guarantees for \CSLPD\ based sparse recovery problems.

Let us mention that in~\cite{GIKS_RIP, XH_expander} expansion arguments were
used to directly obtain similar types of performance guarantees for compressed
sensing; the comparison of these guarantees to the guarantees that can be
obtained through our channel-coding-based arguments remains as future work.

\subsection{The Role of Binary-Input Channels Beyond the BSC
                     for \CSLPD}
\label{sec:beyond:BSC:1}

In Lemma~\ref{lemma:from:bsc:to:cs:lpd:1} we made a connection between
performance guarantees for the BSC under \CCLPD\ on the one hand and the
strict nullspace property $\SNSPR(k, C)$ for $C = 1$ on the other hand. In
this subsection we want to mention that one can establish a connection between
performance guarantees for a certain class of binary-input channels under
\CSLPD\ and the strict nullspace property $\SNSPR(k, C)$ for $C > 1$. This
class of channels consists of binary-input memoryless channels where for all
output symbols the magnitude of the log-likelihood ratio is bounded by some
constant $W \in \Rpp$. Without going into the details, the results
from~\cite{Feldman:Koetter:Vontobel:05:1} (which generalize results
from~\cite{Feldman:Malkin:Servedio:Stein:Wainwright:07:1}) can be used to
establish this connection.\footnote{Note that
  in~\cite{Feldman:Koetter:Vontobel:05:1}, ``This suggests that the asymptotic
  advantage over [\ldots] is gained not by quantization, but rather by
  restricting the LLRs to have finite support.''  should read ``This suggests
  that the asymptotic advantage over [\ldots] is gained not by quantization,
  but rather by restricting the LLRs to have bounded support.''}

The results of this section will be discussed in more detail in a longer
version of the present paper.

\subsection{Connection between AWGNC Pseudo-Weight and
                    $\ell_2 / \ell_1$ Guarantees}

\begin{theorem}
  Let $\matrHCS \in \{ 0, 1 \}^{m \times n}$ be a measurement matrix and
  let $\vs$ and $\ve$ be such that $\vs = \matrHCS \cdot \ve$. Moreover, let
  $\setS \subseteq \setI(\matrHCS)$ with $\card{\setS} = k$, and let $C'$ be
  an arbitrary positive real number with $C' > 4k$. Then the estimate $\hve$
  produced by \CSLPD\ will satisfy
  \begin{align*}
    \twonorm{\ve - \hve}
      &\leq \frac{C''}{\sqrt{k}}
            \cdot
            \onenorm{\veSc}
    \qquad
    \text{with}
    \qquad
    C''
      \defeq
        \frac{1}{\sqrt{\frac{C'}{4k}} - 1},
  \end{align*}
  if $\wpsAWGNC(|\vnnu|) \geq C'$ holds for all $\vnnu \in \nullspaceR(\matrHCS)
  \setminus \{ \vzero \}$. (In particular, this latter condition is satisfied
  for a measurement matrix $\matrHCS$ with $\wpsAWGNCmin(\matrHCS) \geq
  C'$.)
\end{theorem}

\begin{proof}
  By definition, $\ve$ is the original signal. Since $\matrHCS \cdot \ve =
  \vs$ and $\matrHCS \cdot \hve = \vs$, it easily follows that $\vnnu \defeq \ve
  - \hve$ is in the nullspace of $\matrHCS$. So,
  \begin{align}
    \!\!\!\!\!
    \onenorm{\veS}
    +
    \onenorm{\veSc}
      &= \onenorm{\ve} \nonumber \\
      &\overset{\text{(a)}}{\geq}
         \onenorm{\hve} \nonumber \\
      &= \onenorm{\ve + \vnnu} \nonumber \\
      &= \onenorm{\veS + \vnnuS}
         + 
         \onenorm{\veSc + \vnnuSc} \nonumber \\
      &\overset{\text{(b)}}{\geq}
         \onenorm{\veS}
         -
         \onenorm{\vnnuS}
         +
         \onenorm{\vnnuSc}
         -
         \onenorm{\veSc} \nonumber \\
      &\overset{\text{(c)}}{\geq}
         \onenorm{\veS}
         \!+\!
         \left(
           \!
           \sqrt{C'}
           \!-\!
           2\sqrt{k}
         \right) \!
         \twonorm{\vnnu}
         \!-\!
         \onenorm{\veSc},
           \label{eq:ell2:ell1:guarantee:1}
  \end{align}
  where step~$\text{(a)}$ follows from the fact that the solution to \CSLPD\
  satisfies $\onenorm{\hve} \leq \onenorm{\ve}$ and where step~$\text{(b)}$
  follows from applying the triangle inequality for the $\ell_1$ norm
  twice. Moreover, step~$\text{(c)}$ follows from
  \begin{align*}
    -
    \onenorm{\vnnuS}
    +
    \onenorm{\vnnuSc}
      &= \onenorm{\vnnu}
         -
         2
           \onenorm{\vnnuS} \\
      &\overset{\text{(d)}}{\geq}
         \sqrt{C'}
           \twonorm{\vnnu}
         -
         2
           \onenorm{\vnnuS} \\
      &\overset{\text{(e)}}{\geq}
         \sqrt{C'}
           \twonorm{\vnnu}
         -
         2
           \sqrt{k}
           \twonorm{\vnnuS} \\
      &\overset{\text{(f)}}{\geq}
         \sqrt{C'}
           \twonorm{\vnnu}
         -
         2
           \sqrt{k}
           \twonorm{\vnnu} \\
      &= \left(
           \sqrt{C'}
           -
           2\sqrt{k}
         \right)
         \twonorm{\vnnu},
  \end{align*}
  where step~$\text{(d)}$ follows from the assumption that $\wpsAWGNC(|\vnnu|)
  \geq C'$ for all $\vnnu \in \nullspaceR(\matrHCS) \setminus \{ \vzero \}$,
  i.e., $\onenorm{\vnnu} \geq \sqrt{C'} \cdot \twonorm{\vnnu}$ for all $\vnnu
  \in \nullspaceR(\matrHCS)$, where step~$\text{(e)}$ follows from the
  inequality $\onenorm{\va} \leq \sqrt{k} \cdot \twonorm{\va}$ that holds for
  any real vector $\va$ of length $k$, and where step~$\text{(f)}$ follows the
  inequality $\twonorm{\va_{\set{S}}} \leq \twonorm{\va}$ that holds for any
  real vector $\va$ whose set of coordinate indices includes
  $\set{S}$. Subtracting the term $\onenorm{\veS}$ on both sides
  of~\eqref{eq:ell2:ell1:guarantee:1}, and solving for $\twonorm{\vnnu} =
  \twonorm{\ve - \hve}$ yields the promised result.
\end{proof}

\subsection{Connection between Max-Fractional Weight and
                    $\ell_{\infty}/ \ell_1$ Guarantees}

\begin{theorem}
  Let $\matrHCS \in \{ 0, 1 \}^{m \times n}$ be a measurement matrix and
  let $\vs$ and $\ve$ be such that $\vs = \matrHCS \cdot \ve$. Moreover, let
  $\setS \subseteq \setI(\matrHCS)$ with $\card{\setS} = k$, and let $C'$ be
  an arbitrary positive real number with $C' > 2k$. Then the estimate $\hve$
  produced by \CSLPD\ will satisfy
  \begin{align*}
    \infnorm{\ve - \hve}
      &\leq \frac{C''}{k}
            \cdot
            \onenorm{\veSc}
    \qquad
    \text{with}
    \qquad
    C''
      \defeq
        \frac{1}{\frac{C'}{2k} - 1},
  \end{align*}
  if $\wmaxfr(|\vnnu|) {\geq} C'$ holds for all $\vnnu \in \nullspaceR(\matrHCS)
  \setminus \{ \vzero \}$. (In particular, this latter condition is satisfied
  for a measurement matrix $\matrHCS$ with $\wmaxfrmin(\matrHCS) \geq C'$.)
\end{theorem}

\begin{proof}
  By definition, $\ve$ is the original signal. Since $\matrHCS
  \cdot \ve = \vs$ and $\matrHCS \cdot \hve = \vs$, it easily follows that
  $\vnnu \defeq \ve - \hve$ is in the nullspace of $\matrHCS$. So,
  \begin{align}
    \!\!\!\!
    \onenorm{\veS}
    +
    \onenorm{\veSc}
      &= \onenorm{\ve} \nonumber \\
      &\overset{\text{(a)}}{\geq}
         \onenorm{\hve} \nonumber \\
      &= \onenorm{\ve + \vnnu} \nonumber \\
      &= \onenorm{\veS + \vnnuS}
         + 
         \onenorm{\veSc + \vnnuSc} \nonumber \\
      &\overset{\text{(b)}}{\geq}
         \onenorm{\veS}
         -
         \onenorm{\vnnuS}
         +
         \onenorm{\vnnuSc}
         -
         \onenorm{\veSc} \nonumber \\
      &\overset{\text{(c)}}{\geq}
         \onenorm{\veS}
         +
         \left(
           C'
           -
           2k
         \right)
         \cdot
         \infnorm{\vnnu}
         -
         \onenorm{\veSc},
           \label{eq:ellinfty:ell1:guarantee:1}
  \end{align}
  where step~$\text{(a)}$ follows from the fact that the solution to \CSLPD\
  satisfies $\onenorm{\hve} \leq \onenorm{\ve}$ and where step~$\text{(b)}$
  follows from applying the triangle inequality for the $\ell_1$ norm
  twice. Moreover, step~$\text{(c)}$ follows from
  \begin{align*}
    -
    \onenorm{\vnnuS}
    +
    \onenorm{\vnnuSc}
      &= \onenorm{\vnnu}
         -
         2
           \cdot
           \onenorm{\vnnuS} \\
      &\overset{\text{(d)}}{\geq}
         C'
           \cdot
           \infnorm{\vnnu}
         \!-\!
         2
           \cdot
           \onenorm{\vnnuS} \\
      &\overset{\text{(e)}}{\geq}
         C'
           \cdot
           \infnorm{\vnnu}
         \!-\!
         2k
           \cdot
           \infnorm{\vnnuS} \\
      &\overset{\text{(f)}}{\geq}
         \sqrt{C'}
           \cdot
           \infnorm{\vnnu}
         -
         2k
           \cdot
           \infnorm{\vnnu} \\
      &= \left(
           C'
           -
           2k
         \right)
         \cdot
         \infnorm{\vnnu},
  \end{align*}
  where step~$\text{(d)}$ follows from the assumption that $\wmaxfr(|\vnnu|)
  \geq C'$ for all $\vnnu \in \nullspaceR(\matrHCS) \setminus \{ \vzero \}$,
  i.e., $\onenorm{\vnnu} \geq C' \cdot \infnorm{\vnnu}$ for all $\vnnu \in
  \nullspaceR(\matrHCS)$, where step $\text{(e)}$ follows from the inequality
  $\onenorm{\va} \leq k \cdot \infnorm{\va}$ that holds for any real vector
  $\va$ of length $k$, and where step~$\text{(f)}$ follows the inequality
  $\infnorm{\va_{\set{S}}} \leq \infnorm{\va}$ that holds for any real vector
  $\va$ whose set of coordinate indices includes $\set{S}$. Subtracting the
  term $\onenorm{\veS}$ on both sides of~\eqref{eq:ellinfty:ell1:guarantee:1},
  and solving for $\infnorm{\vnnu} = \infnorm{\ve - \hve}$ yields the promised
  result.
\end{proof}

\subsection{Connection between BEC Pseudo-Weight and \CSLPD}

For the binary erasure channel, \CCLPD\ is identical to the peeling
decoder~\cite{Richardson:Urbanke:08:1} that is just solving a system of linear
equations by only using back-substitution. We can define an analogous
compressed sensing problem by assuming that the compressed sensing decoder is
\emph{given the support} of the sparse signal $\ve$ and decoding simply
involves trying to recover the values of the non-zero entries by
back-substitution, similarly to iterative matching pursuit. In this case it is
clear that \CCLPD\ for the BEC and the described compressed sensing decoder
have identical performance since back-substitution behaves exactly the same
way over any field, be it the field of real numbers or any finite field.
(Note that whereas the result of the \CCLPD\ for the BEC equals the result of
the back-substitution-based decoder for the BEC, the same is not true for
compressed sensing, i.e., \CSLPD\ with given support of the sparse signal can
be strictly better than the back-substitution-based decoder with given support
of the sparse signal.)

\section{Conclusions and Future work}
\label{sec:conclusions:1}

Based on the observation that points in the nullspace of a zero-one matrix
(considered as a real measurement matrix) can be mapped to points in the
fundamental cone of the same matrix (considered as the parity-check matrix of
a code over $\GF{2}$), we were able to establish a connection between \CSLPD\
and \CCLPD.

In addition to \CSLPD, a number of combinatorial algorithms
(e.g.~\cite{DM_SP09, XH_expander, Tropp_OMP, GIKS_RIP, GLW_random}) have been
proposed for compressed sensing problems, with the benefit of faster decoding
complexity and comparable performance to \CSLPD.  It would be interesting to
investigate if the connection of sparse recovery problems to channel coding
extends in a similar manner for these decoders. One example of such a clear
connection is the bit-flipping algorithm of Sipser and
Spielman~\cite{Sipser:Spielman:96} and the corresponding algorithm for
compressed sensing by Xu and Hassibi~\cite{XH_expander}. Connections of
message-passing decoders for compressed sensing problems were also recently
discussed in~\cite{ZhangPfister_09}.

Other interesting directions involve using optimized channel coding matrices
with randomized or deterministic
constructions~(e.g., see~\cite{Richardson:Urbanke:08:1}) to create measurement
matrices. Another is using ideas for improving the performance of a given
measurement matrix (for example by removing short cycles), with possible
theoretical guarantees. Finally, one interesting question relates to being able
to certify in polynomial time that a given measurement matrix has good
performance.

In any case, we hope that the connection between \CSLPD\ and \CCLPD\ that was
discussed in this paper will help deepen the understanding of the role of
linear programming relaxations for sparse recovery and for channel coding, in
particular by translating results from one field to the other.

\section*{Acknowledgments}

The first author would like to thank Prof.\ Babak Hassibi for stimulating
discussions that helped in the development of this research.

\bibliographystyle{IEEEtran}

\bibliography{allerton2009_references}

\end{document}